\documentclass[epjCONF,columns]{svjour} 
\usepackage{graphics}
\usepackage[varg]{txfonts} % Times fonts
\usepackage[latin1]{inputenc}
\session-title{Hadron Collider Physics Symposium 2011, Paris}
\newcommand{\met}{\makebox[2.4ex]{\ensuremath{\not\!\!\!\; E_{\textrm{T}}}}}

\makeatletter
\DeclareRobustCommand*{\bfseries}{%
  \not@math@alphabet\bfseries\mathbf
  \fontseries\bfdefault\selectfont
  \boldmath
}
\makeatother

\begin{document}
\title{Measurement of the Charge Asymmetry in Top Quark Pair Production}
\author{Christian B\"oser \inst{1} on behalf of the CMS collaboration}
\institute{Institut f\"ur Experimentelle Kernphysik, Karlsruhe Institute of Technology (KIT), 76128 Karlsruhe, Germany}
\abstract{
We present a measurement of the charge asymmetry in top quark pair production using an integrated luminosity of 1.09 fb$^{-1}$ collected with the CMS detector. Top quark pairs with a signature of one electron or muon and four or more jets, at least one of them b tagged, are selected. At the LHC a small charge asymmetry in the rapidity distributions of top and antitop quarks is predicted. Therein slightly broader rapidity distributions for top quarks are expected, while antitop quarks are produced more centrally and possess narrower rapidity distributions. We determine the charge asymmetry based on two different sensitive variables and the results are compared with the most precise standard model theory predictions using a dedicated unfolding technique. 
} %end of abstract
\maketitle
\section{Introduction}
\label{sec:intro}
In the Standard Model (SM), a small charge asymmetry in $\rm t\bar{t}$ production via quark-antiquark annihilation is predicted. It can be explained via the interference between the Born and the box diagram, and between initial- and final-state radiation, which link the flight directions of the top quarks and antiquarks to the directions of the initial quarks and antiquarks, respectively \cite{Kuehn:1998,Kuehn:1999}. At the Tevatron proton-antiproton collider this effect leads to a forward-backward asymmetry which is visible in the difference in rapidity $y$ of top quarks and antiquarks. Recent measurements by the CDF and D\O\ collaborations \cite{CDF-1,D0-1} measure asymmetries which are about two standard deviations larger than the predicted SM value of about 0.08 \cite{Kuehn:2011}. The CDF collaboration reports an even larger asymmetry compared to the SM prediction at high $\rm t\bar{t}$ invariant masses \cite{CDF-2}. These results have led to speculations that the large asymmetry might be generated by additional axial couplings of the gluon or by heavy particles with unequal vector and axial-vector couplings to top quarks and antiquarks.

In proton-proton collisions at the LHC the charge asymmetry leads to an central-peripheral asymmetry, which can be observed through the difference in the absolute values of the pseudorapidities of top quarks and antiquarks $\Delta |\eta|=|\eta_t|-|\eta_{\bar{t}}|$. We also measure the charge asymmetry in the observable used by the Tevatron experiments multiplied by a factor that accounts for the boost of the $\rm t\bar{t}$ system, yielding $\Delta y^2 = (y_t - y_{\bar{t}})\cdot(y_t + y_{\bar{t}})=(y_t^2 - y_{\bar{t}}^2)$, motivated in \cite{y2}. Using either of the two variables, the charge asymmetry can be defined as 
\begin{equation}
\label{eq:ACDC}
A_C = \frac{N^+ - N^-}{N^+ + N^-}\textrm{\,,}
\end{equation}  
where $N^+$ and $N^-$ are the number of events with positive and negative values in the sensitive variable, respectively. The SM prediction for $\Delta |\eta|$ is $A_C^\eta (\textrm{theory})= 0.0136\pm 0.0008$ \cite{Kuehn:2011} and for $\Delta y^2$ it is $A_C^y(\textrm{theory}) = 0.0115\pm 0.0006$ \cite{Kuehn:2011}, where the uncertainties arise from scale variations and particle density function uncertainties. These values are much smaller than the SM predictions for the Tevatron, because of the large fraction of gluon-gluon induced $\rm t\bar{t}$ production. 

\section{The CMS Detector}
\label{sec:Detector}
The central feature of the Compact Muon Solenoid (CMS) apparatus is a superconducting solenoid, of 6~m internal diameter, providing a field of 3.8~T. Within the field volume are the silicon pixel and strip tracker, the crystal electromagnetic calorimeter (ECAL) and the brass/scintillator hadron calorimeter (HCAL). Muons are measured in gas-ionization detectors embedded in the steel return yoke. In addition to the barrel and endcap detectors, CMS has extensive forward calorimetry. A much more detailed description of CMS can be found elsewhere~\cite{JINST}. 

\section{Selection of Events}
\label{sec:selection}
The measurement is performed in the electron$+$jets and muon$+$jets decay channels, using proton-proton collision events at a center of mass energy of 7 TeV recorded with the CMS detector up to July 2011. The amount of data corresponds to an integrated luminosity of $1.09~\pm 0.07\,\mathrm{fb}^{-1}$. Only events selected by special trigger algorithms which search for electrons or muons together with at least three jets with at least $30\,\mathrm{GeV}/c$ of transverse momentum are used. In addition, the presence of a good primary vertex (PV) is required. 

In the electron$+$jets channel electron candidates are required to possess a transverse energy larger than 30 GeV and to lie in the region $|\eta| < 2.5$, whereas in the muon$+$jets channel muon candidates must have a transverse momentum greater than $20\,\mathrm{GeV}/c$ and must lie within the muon trigger acceptance ($|\eta |<2.1$). Both lepton candidates have to be isolated and must pass additional quality criteria. Jets are required to have a corrected jet energy $p_{\rm{T}} > 30\,\mathrm{GeV}/c$ and $|\eta |<$~2.4. 

Only events with exactly one isolated charged lepton and at least four jets, where one of them is tagged as a \mbox{$b$ jet}, are selected. In both channels events with additional more loosely defined charged leptons are discarded.

Applying the event selections to the recorded dataset of $1.09\,\mathrm{fb}^{-1}$ in total 12757 events are selected, 5665 in the electron$+$jets channel and 7092 in the muon$+$jets channel.

\section{Background Estimation}
\label{sec:background}
The measurement of the charge asymmetry requires a precise knowledge of the amount if background events in the analyzed dataset. Therefore we apply a data-driven background estimation using two kinematic variables, the missing transverse energy and M3, which is the invariant mass of the three jets with the largest vectorially summed transverse momentum. The selected datasets in the electron$+$jets and muon$+$jets channel are separated each in two subsets, events with \mbox{$\met~ < 40\,\mathrm{GeV}$} and events with \mbox{$\met~ > 40\,\mathrm{GeV}$}. We fit the \met~ distribution in the low-\met~ dataset and the M3 distribution in the high-\met~ dataset simultaneously. The background estimation is performed individually for the two lepton types. $\rm{W}+$jets contributions from $\rm{W}^{+}$ and $\rm{W}^{-}$ are estimated individually. In order to simulate the QCD background properly a QCD model from data is used by inverting the cut on the relative isolation. The fits are performed as binned likelihood fits using the \texttt{theta} framework~\cite{theta} yielding a signal purity of about 80 \%. 

\section{Reconstruction of $\rm{t\bar{t}}$ Pairs}
\label{sec:rec}
In order to fully reconstruct the top quarks in the combined lepton$+$jets decay channel, the selected physical objects are consecutively assigned to final state particles. The assignment is not unambiguous and we define a criterion $\psi$ based only on reconstructible quantities in order to give each possible hypothesis a probability to be the best hypothesis for the event. For $\psi$ the three linearly decorrelated masses $m_1$, $m_2$, and $m_3$ arising from the masses of the reconstructed top quarks and the hadronically decaying $W$ boson, as well as the $b$ tagger discriminator values for the jets assigned to the two $b$ quarks and to the two light quarks are taken into account. The hypothesis with the biggest value of $\psi$ is then chosen in each event for further consideration. 

\section{Measurement of the $\rm t\bar{t}$ Charge Asymmetry}
\label{sec:measurement}
\begin{figure}
\resizebox{0.75\columnwidth}{!}{\includegraphics{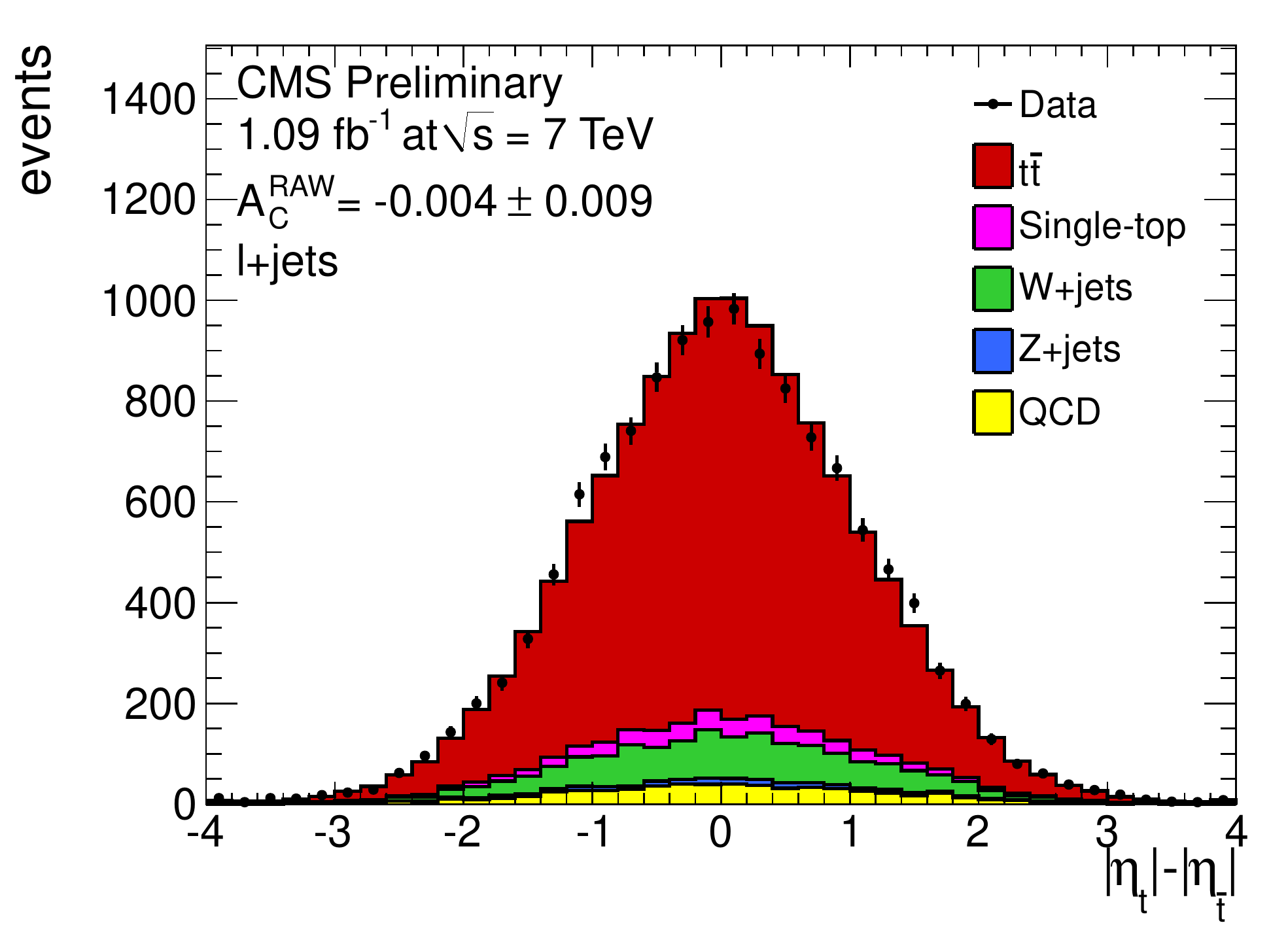}}
\caption{Reconstructed $\Delta |\eta|$ distribution for the combined lepton$+$jets channel.}
\label{fig:Asy_DataMC}      
\end{figure}
In figure~\ref{fig:Asy_DataMC} the distribution of $\Delta |\eta|$ in the lepton$+$jets channel obtained from the reconstructed top and antitop quarks is shown. Both reconstructed distributions ($\Delta |\eta|$ and $\Delta y^2$) can be used to calculate an uncorrected charge asymmetry $A_{C}^{RAW}$ by simply using the definition in equation~\ref{eq:ACDC} yielding \mbox{$A_{C}^{RAW}(\eta) = -0.004 \pm 0.009$} and \mbox{$A_{C}^{RAW}(y) = -0.004 \pm 0.009$}.

These values are not directly comparable with any theoretically motivated prediction, since several effects bias the measurement at this stage. First of all, 20\% of events used to measure $A_{C}^{RAW}$ arise from background processes. Therefore we subtract the predicted amount of background events from the measured distributions. In this step a Gaussian error propagation is performed taking the measured uncertainties on the background rates as input.

An even larger effect is due to the event selection efficiency and to the imperfect reconstruction method. Depending on the true value of $\Delta|\eta|$ (or $\Delta y^{2}$) the probability for an event to survive the event selection varies, which leads to distortions in the reconstructed distributions. Furthermore, in many events not exactly the best possible jet parton assignment and neutrino reconstruction is chosen. But even when the best possible reconstruction hypothesis is chosen, this does not guarantee that the top quark four-momenta are reconstructed correctly. Due to the finite energy resolution of the calorimeters and the jet reconstruction, and due to the possibility that jets from the $\rm t\bar{t}$ decay can lie outside the detector acceptance, a perfect reconstruction is a priori not possible.
In order to investigate all these effects one can analyze simulated $\rm t\bar{t}$ events, where the four momenta of the generated top quarks are compared to the four vectors reconstructed from measured objects like jets and leptons. 
From these studies one finds that the distortion of the distributions can be factorized into effects due to the event selection efficiency and due to the event reconstruction procedure.

To correct for these effects, we apply a regularized unfolding procedure~\cite{Blobel:2002pu}. The unfolding algorithm corrects the measured spectrum for migration and efficiency effects by applying a generalized matrix inversion method.
The measured spectrum is divided into 12 bins, where the different widths of the bins have been chosen such that they contain approximately equal numbers of events. For the unfolded spectrum six bins have been used. To regularize the problem and to avoid unphysical fluctuations two additional terms, a regularization term and a normalization term, are used~ \cite{Tikhonov,Phillips}.

The performance of the unfolding algorithm is tested in 50,000 pseudo experiments. For each pseudo experiment a pseudo dataset is created and unfolded with the method described above. In each pseudo experiment the unfolded spectrum is compared with the generated distribution and a very good agreement is found putting the analysis on solid ground. Also with pseudo experiments in which we re-weight the events  in order to generate a non-zero asymmetry between $-0.21$ and $+0.23$ a linear behavior of the average fit results as a function of the true values is found. While for the $\Delta |\eta|$ variable the agreement is almost perfect, for $\Delta y^{2}$ a small bias in the order of $0.1\%$ has been observed for which the measured asymmetry has to be corrected by a factor of 0.94. 

The measurement of the charge asymmetry $A_{C}$ might be affected by several systematic sources uncertainties. Only systematic uncertainties influencing the direction of the reconstructed top quark momenta can change the value of the reconstructed charge asymmetry. The overall selection efficiency and acceptance will not change the measured values. For each source of systematic pseudo datasets are drawn from systematically shifted samples, to which the unfolding method with the standard templates are applied. The largest systematic uncertainties arise from the variation of the $Q^2$ scale and matching threshold and from the variation of initial- and final-state-radiation in the used $\rm t\bar{t}$ signal simulated sample.

\begin{figure}
\resizebox{0.75\columnwidth}{!}{\includegraphics{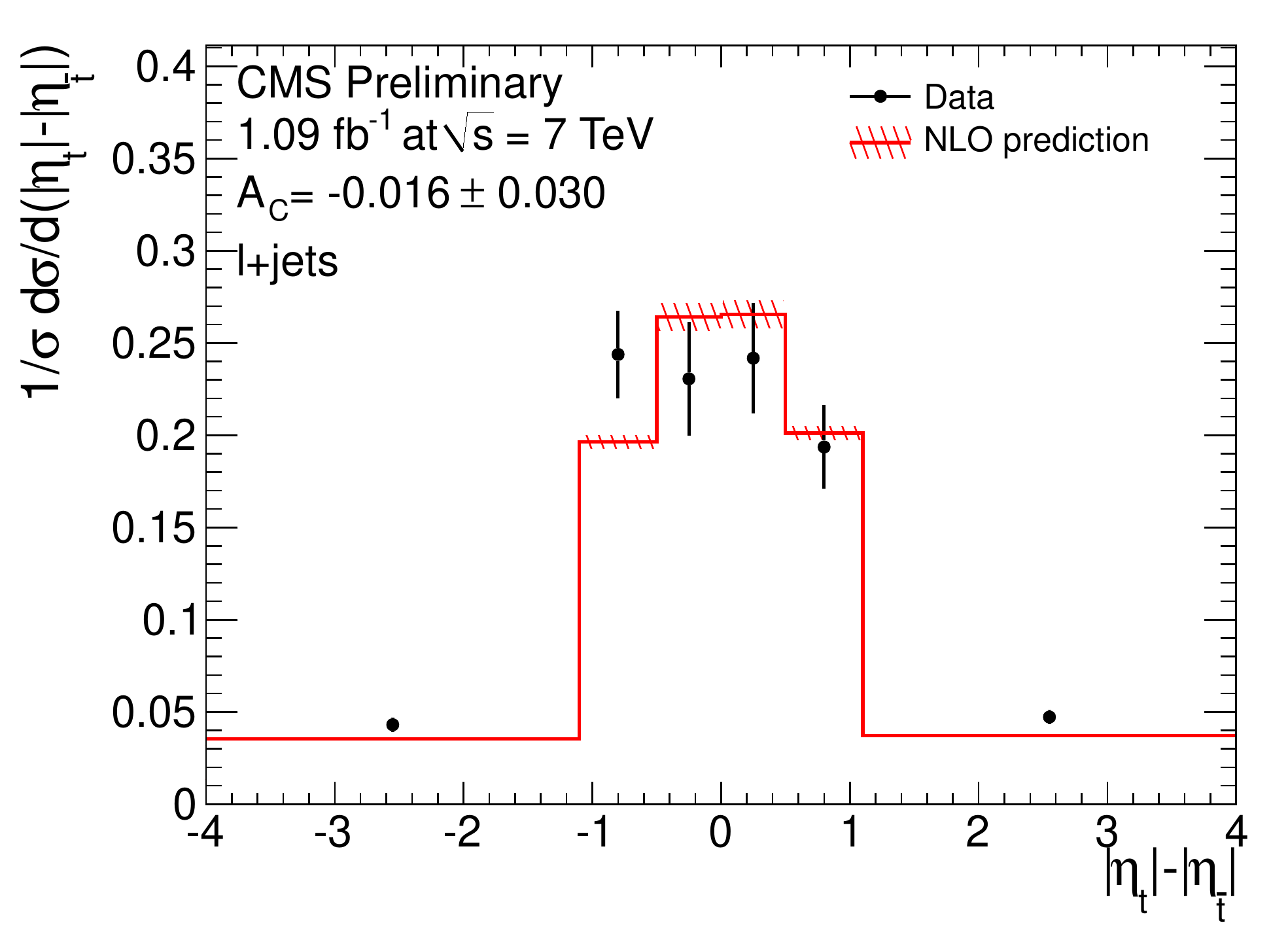}}
\caption{Unfolded $\Delta |\eta|$ normalized spectrum. The NLO prediction is based on the calculation of Ref. \cite{Kuehn:2011}.}
\label{fig:Asy_Results}      
\end{figure}

We apply the described unfolding procedure to the measured $\Delta|\eta|$ distribution as well as to the distribution of the second variable, $\Delta y^{2}$. Figure~\ref{fig:Asy_Results} shows the unfolded spectrum for $\Delta|\eta|$ used for computing the asymmetries together with the SM prediction at NLO. In the unfolded $\Delta|\eta|$ distribution we measure an asymmetry of 
\begin{equation}
A_{C}^{\eta} = -0.016 \pm 0.030~(\mathrm{stat.})^{+0.010}_{-0.019}(\mathrm{syst.})~,
\end{equation}
while in $\Delta y^{2}$ we measure an unfolded and corrected (divided by 0.94) asymmetry of 

\begin{equation}
A_{C}^{y} = -0.013 \pm 0.026~(\mathrm{stat.})^{+0.026}_{-0.021}(\mathrm{syst.})~.
\end{equation}

Both measured values are within the uncertainties in agreement with the theory predictions. 

\begin{figure}
\resizebox{0.75\columnwidth}{!}{\includegraphics{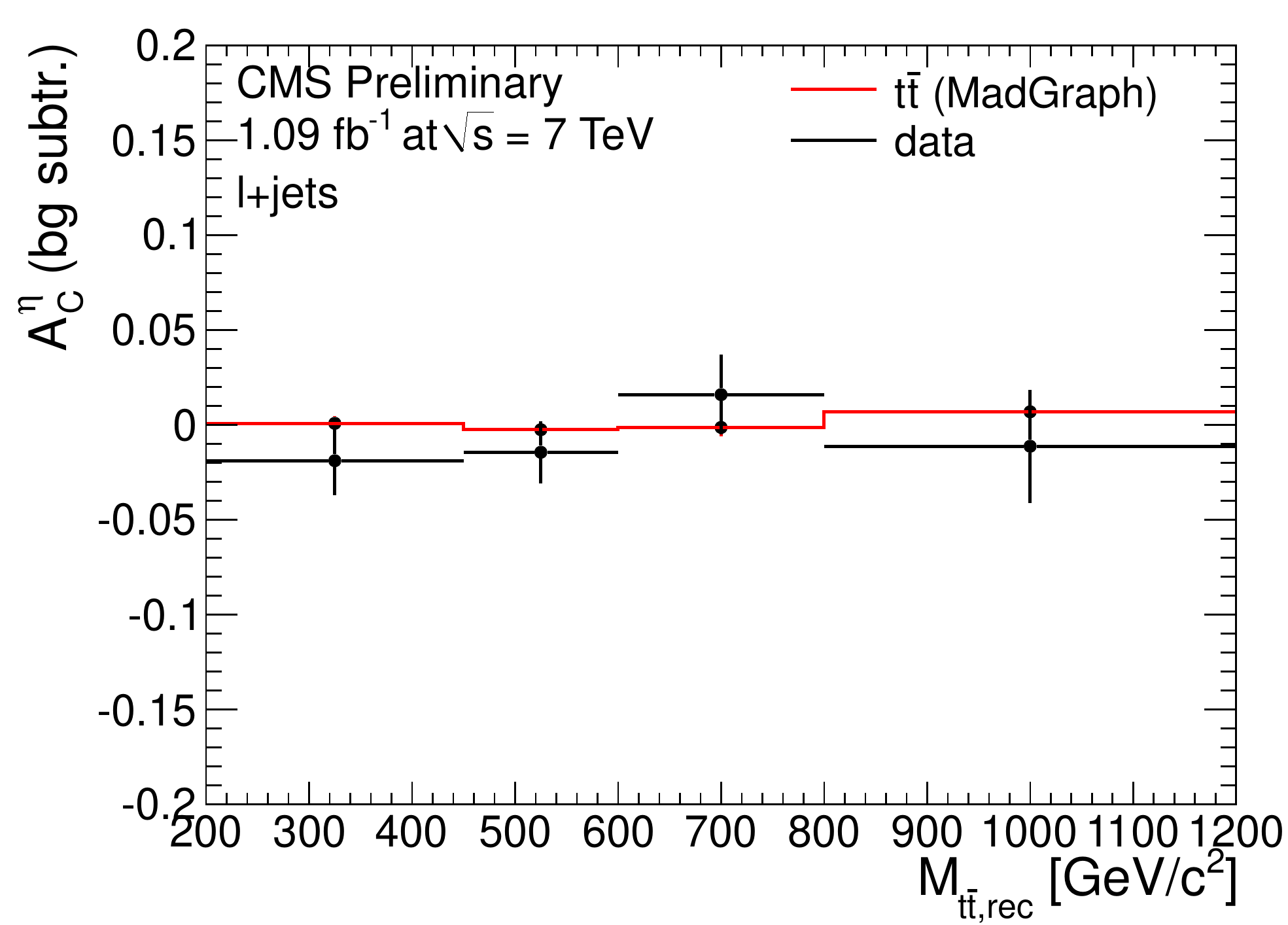}}
\caption{Background-subtracted asymmetries for $\Delta |\eta|$ as function of the reconstructed $\rm t\bar{t}$ invariant masses.}
\label{fig:Mttbar}      
\end{figure}

One can also measure the background subtracted asymmetry as a function of the reconstructed invariant mass of the $\rm{t\bar{t}}$ system to investigate whether one can see a dependence of the asymmetry on $m_{\rm{t\bar{t}}}$. Figure~\ref{fig:Mttbar} shows the result for the $\Delta |\eta|$ variable, where no increase of the asymmetry for increasing $m_{\rm{t\bar{t}}}$ can be seen. The same behavior is found for $\Delta y^2$. However, these studies allow only for a qualitative statement, while for a quantitative statement a proper simultaneous unfolding in the sensitive variable as well as in the invariant mass of the $\rm{t\bar{t}}$ system has to be performed.

\section{Conclusion}
\label{sec:conclusion}
The analysis \cite{CMS-PAS-TOP-11-014} presented here provides the measurement of the charge asymmetry in $\rm t\bar{t}$ events using a dataset corresponding to an integrated luminosity of $1.09\,\mathrm{fb}^{-1}$. A selection of $\rm t\bar{t}$ events has been performed in the lepton$+$jets decay channel and the amount of background events in the selected dataset has been estimated via a binned likelihood fit. Thereafter the four-momenta of the top quarks have been reconstructed and corrected using a regularized unfolding method. As opposed to the Tevatron results, we measure a small negative asymmetry which is within the uncertainties still very well compatible with the SM predictions and shows no tendency to large deviations from the prediction.
Also the background subtracted asymmetry as a function of the invariant mass of the $\rm{t\bar{t}}$ system shows no tendency to larger values for large invariant masses. For a quantitative statement of this behavior a proper simultaneous unfolding in both, the sensitive variable and the invariant mass of the $\rm{t\bar{t}}$ system, will be performed in the future in order to confirm or to refute the Tevatron findings.


\begin{thebibliography}{}
\bibitem{Kuehn:1998} J. H. K\"uhn and G. Rodrigo, ``Charge Asymmetry in Hadroproduction of Heavy Quarks'', Phys. Rev. Lett. \textbf{81}, (1998) 49
\bibitem{Kuehn:1999} J. H. K\"uhn and G. Rodrigo, ``Charge Asymmetry of Heavy Quarks at Hadron Colliders'', Phys. Rev. D \textbf{59} (1999) 054017
\bibitem{CDF-1} The CDF collaboration, ``Combination of the Forward-Backward Asymmetry in the Top Pair Production from L+J and DIL Channels using 5 fb$^{-1}$'', CDF Note 10584 (2011)
\bibitem{D0-1} The D\O\ collaboration, ``Forward-Backward Asymmetry in Top Quark-Antiquark Production'', (2011), arXiv:1007.4995
\bibitem{Kuehn:2011} J. H. K\"uhn and G. Rodrigo, ``Charge Asymmetries of Top Quarks at Hadron Colliders Revisited'', (2011), arXiv:1109.6830
\bibitem{CDF-2} The CDF collaboration, ``Evidence for a Mass Dependent Forward-Backward Asymmetry in Top Quark Pair Production'', Phys. Rev. D \textbf{83}, (2011) 112003
\bibitem{y2} S. Jung, A. Pierce and J. D. Wells, ``Top Quark Asymmetry from a Non-Abelian horizontal Symmetry'', arXiv:1003.4835 
\bibitem{JINST} The CMS collaboration, ``The CMS experiment at the CERN LHC'', JINST 0803:S08004, 2008.
\bibitem{theta} J. Ott, \texttt{www.theta-framework.org}, (2011)
\bibitem{Blobel:2002pu} V.Blobel, ``An unfolding method for high energy physics experiments'', (2002), arXiv:hep-ex/0208022
\bibitem{Tikhonov} A. Tikhonov, ``Solution of incorrectly formulated problems and the regularization method'', Soviet Mathematics Doklady  \textbf{4}, (1963) 1035--1038
\bibitem{Phillips} D. L. Phillips, `` A Technique for the Numerical Solution of Certain Integral Equations of the First Kind'', J. ACM \textbf{9}, (1962) 84--97
\bibitem{CMS-PAS-TOP-11-014} The CMS collaboration, ``Measurement of the Charge Asymmetry in Top Quark Pair Production'', CMS Physics Analysis Summary TOP-11-014, (2011)


\end{thebibliography}
\end{document}